\documentclass[twocolumn]{revtex4}
\usepackage[dvips]{graphicx}
\usepackage{bm}
\usepackage{amsmath}
\begin{document}
\title{%
Final-state read-out of exciton qubits by observing resonantly excited  photoluminescence in quantum dots
}

\author{K.~Kuroda}
\author{T.~Kuroda}
\altaffiliation{%
Author to whom correspondence should be addressed; 
also at: PRESTO, Japan Science and Technology Agency.}
\author{K.~Watanabe}
\altaffiliation[Present address: ]{%
Institute of Industrial Science, The University of Tokyo, 
Tokyo 153-8505, Japan.}
\author{T.~Mano}
\author{K.~Sakoda}
\altaffiliation[Also at: ]{%
Graduate School of Pure and Applied Sciences, University of Tsukuba,
1-1-1 Tennodai, Tsukuba 305-8577, Japan.}
\author{G.~Kido}
\author{N.~Koguchi}
\affiliation{%
Quantum Dot Research Center, National Institute for Materials Science, 
1-1 Namiki, Tsukuba 305-0044, Japan.}
\date{\today}
\begin{abstract}
We report on a new approach to detect excitonic qubits in semiconductor quantum dots by observing spontaneous emissions from the relevant qubit level. The ground state of excitons is resonantly excited by picosecond optical pulses. Emissions from the same state are temporally resolved with picosecond time resolution. To capture weak emissions, we greatly suppress the elastic scattering of excitation beams, by applying obliquely incident geometry to the micro photoluminescence set-up. Rabi oscillations of the ground-state excitons appear to be involved in the dependence of emission intensity on excitation amplitude. 

\end{abstract}
\maketitle

Precise detection of a single quantum is a key requirement for establishing quantum information processing in solid states. Single-shot read-out of an individual electron spin has been realized in a mesoscopic quantum dot (QD) using a conductance technique \cite{EHW}. For excitons in self-assembled QDs, the final state after coherent manipulations is determined, in principle, when single-photon spontaneous emissions associated with the exciton annihilation are confirmed. In reality, however, such a straightforward approach has not been followed, because of a small signal of the photon emission: In the framework of exciton-based quantum computations, a qubit is manipulated by intense resonant fields, which obscure a signal emitted by a single QD. 

To avoid this difficulty, several groups have taken an alternative approach to implement a fundamental qubit gate. Refs.~\onlinecite{Kamada,Htoon,Muller} reported on Rabi oscillations of the excited level of excitons, where they measured photoemissions from the ground-state level to read out a final state. Refs.~\onlinecite{Zrenner} and \onlinecite{Stufler} demonstrated very sensitive photocurrent detections for a single QD state. In both cases, however, the relevant qubit had to be surrounded by a fast relaxation/tunneling channel for its high-yield detection. Such a design naturally results in serious reduction in the decoherence times. Moreover, Refs.~\onlinecite{Stievater} and \onlinecite{Li} observed transient absorption of ground-state excitons in a single QD. In this case, the small absorption cross section required them to make modulation-based, long-term averaging, measurements, which are unfavorable for quantum computational applications. Later, Rabi oscillations of ground state excitons were investigated in ensemble QDs by means of nonlinear optical techniques \cite{Borri, Mitsumori}. 

In the present study, we observed spontaneous emissions from ground-state excitons 
after strictly resonant excitation. Great care was taken to reduce the elastic scattering of excitation beams. For this purpose, we irradiated the sample with a nearly-plane-wave field of $\bm{k}_\mathrm{in}$, and collected the signal that was generated into a different mode, $\bm{k}_\mathrm{out}$, where $\bm{k}_\mathrm{in} \neq \bm{k}_\mathrm{out}$. By applying such a non-conventional geometry together with time-resolved detection, we have successfully identified resonant emissions. 
Up to now, resonantly-excited emissions had only been studied with focusing a large number of QDs. \cite{paillard_apl00} The measurement of resonantly-excited emissions 
offers highly sensitive read-out for the qubit final state, since the emission intensity is proportional to the probability of finding an exciton in the QD. As a demonstration of qubit manipulations, we have observed Rabi oscillations, i.e., single qubit rotational gates and their read out. 

The experiments were performed in GaAs self-assembled QDs in an Al$_{0.3}$Ga$_{0.7}$As barrier, grown by the droplet epitaxial technique (Koguchi method) \cite{KTC91, Watanabe1,Watanabe2}. Atomic force microscopy and high-resolution scanning electron microscopy demonstrated the formation of lens-shaped QDs of 16 nm in height and 20 nm in base radius, with a surface density of $1.0\times10^{10}$cm$^{-2}$. Photoluminescence (PL) of the QD ensemble showed a broad spectral band centering at 1.73 eV with 120 meV in full width at half maximum (FWHM). \cite{kuroda_prb02}

\begin{figure}
\includegraphics[angle=-90,width=7cm,clip]{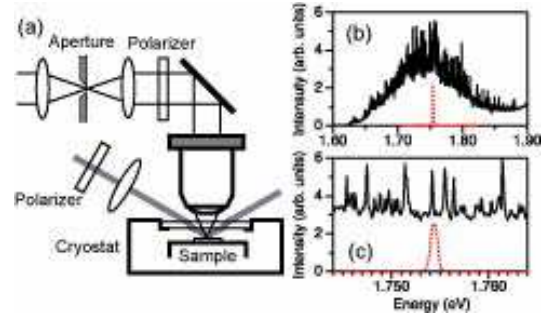}
\caption{(color online) (a) Experimental set-up for the measurement of resonantly-excited single QD photoluminescence: (b) Micro PL spectrum of QD ensembles at off-resonant excitation (solid line) and the spectrum of the lasers for resonant pulsed excitation (dotted line). The expanded view is shown in (c).  }
\label{fig_setup}
\end{figure}

Figure \ref{fig_setup}(a) shows a layout of our set-up. A mode-locked Ti-sapphire laser was used for excitation. It produced picosecond pulses with 0.3~nm FWHM and 76~MHz in repetition rate. The photon energy was chosen to be 1.754 eV, that excited a high energy side of the ensemble PL spectrum -- see Fig.~\ref{fig_setup}(b). The beam was obliquely incident on the sample with the angle of incidence set at $\sim$60 degrees, nearly corresponding to the Brewstar angle of our cryostat window and that of the sample. Application of the \textit{p} polarization to the excitation beams greatly reduced nonresonant elastic scattering. The beam was loosely focused by a convex lens with 100~mm focal length. The excitation spot on the sample was characterized as an elliptic shape of 90 $\mu$m in the major axis and 40 $\mu$m in the minor axis. 

Emissions normal to the surface were collected by an abbreviation-corrected micro objective of N.A. (numerical aperture) $=0.55$. The objective was attached to a 
three-axis translator, to allow for precise movement in the position of detection. Behind the objective lens, a linear polarizer was inserted to reject the colinear polarized component of excitation/scattering lights. Then, the beam was focused on a pinhole acted as spatial filter with a detection spot of $1.0\,\mu$m in diameter. For this configuration, the spot of excitation was much larger than that of detection, so that the QDs were homogeneously irradiated, allowing us to determine excitation density with high accuracy. After passing through the spatially-filtering device, the signal was fed into the entrance slit of a polychromator, and temporally resolved by a synchronously scanning streak camera. Temporal- and spectral resolutions of our set-up were 5~ps and $0.8$~meV, respectively. For spectral analysis, the emission signals were measured by a spectrometer equipped with a cooled charge-coupled device, whose minimal resolution was $0.15$~meV in FWHM. 
During the experiment, the sample was attached to a cold finger of a liquid helium cryostat. All experiments were performed at 4 K. 

We should mention that the present set-up allows us to observe single-QD PL, although the density of our QDs is relatively high. This is due to a small spectral width of our excitation pulses. We estimate the average number of QDs inside the spot of detection to be 80, while the number of excited QDs decreases to $\sim$0.5, according to the narrow spectral band of excitation pulses. In Figs.~\ref{fig_setup}(b) and (c), we show the comparison between the micro PL spectrum of our sample and the spectrum of excitation laser pulses, suggesting a high probability of exciting a single QD. Note that the linewidth of a single QD PL was precisely observed to be $\sim$40 $\mu$eV by means of the Fourier transform technique\cite{kuroda_apl06}. The PL linewidth is, therefore, much smaller than that of excitation pulses. In this case, the laser energy must coincide with the transition energy of a QD, otherwise no QDs would be efficiently excited. To achieve such a resonant condition, we moved the position of detection in a precise manner, until we could capture a highly-luminescent QD, and observe strong resonant PL signals. 

\begin{figure}
\includegraphics[angle=-90,width=7cm,clip]{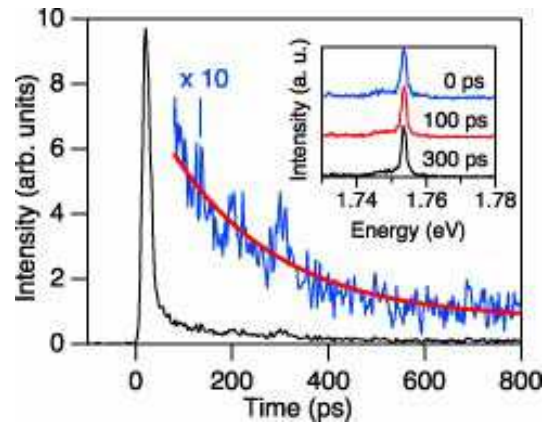}
\caption{(color online) Temporal evolution of PL emissions from a single QD after resonant pulsed excitation. The solid line presents a fit to an exponential function. Transient PL spectra recorded at 0~ps, 100~ps and 300~ps are shown in the inset. The spectra are normalized to their maxima.}
\label{fig_pl}
\end{figure}

Temporal development of PL emissions after pulsed resonant excitation is shown in Fig.~\ref{fig_pl}. An intense signal at 0~ps reflects residual elastic scattering of excitation pulses, caused by the surface roughness of our sample and/or other dielectric materials. Following this instantaneous component, a decaying signal is clearly observed. This is associated with spontaneous emissions of the QD which was resonantly excited. The decay time is estimated to be 220 ($\pm$10)~ps. The spectral line width of the decay signal is found to be smaller than our instrumental response, as shown in the inset of Fig.~\ref{fig_pl}. Moreover, there is no measurable Stokes shift with time. These measurements demonstrate that there is no significant carrier tunneling, and emissions are produced by the QD which has been resonantly excited. 

\begin{figure}
\includegraphics[width=7cm,clip]{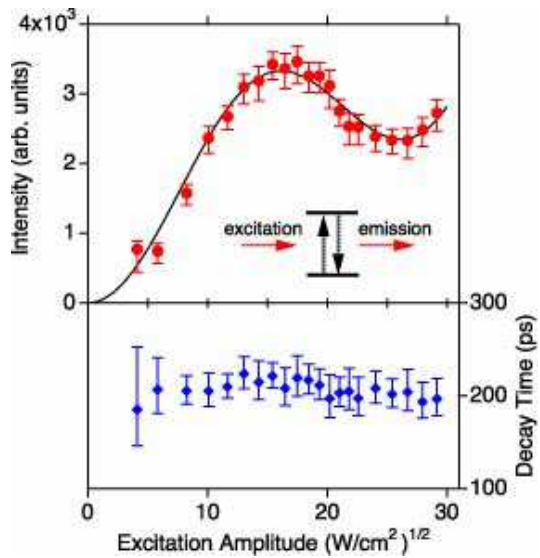} 
\caption{(color online) (top) Dependence of the time-integrated intensity of single QD resonant emissions on excitation amplitude. Fit to the data is shown by the solid line. A schematic view of resonantly-excited emissions is shown in the inset. (bottom) Decay times of the PL transients as a function of excitation amplitude.}
\label{fig_rabi} 
\end{figure}

In Fig.~\ref{fig_rabi} we plot the dependence of the intensity of single QD emissions on excitation density. The emission intensity is evaluated by the temporal integration of the decaying signal after removing the scattering component. Note that the horizontal axis in Fig.~\ref{fig_rabi} is scaled by the square root of excitation density, since it is proportional to the pulse area, that is, the angle for the qubit rotation. An oscillatory feature is clearly found to appear in the density dependence; with increasing excitation density, the signal increases initially, then decreases, and finally increases again, until the excitation density reaches the maximum power available. It is notable that decay times indicate no significant change with varying density, as shown in the lower panel of Fig.~\ref{fig_rabi}. Thus, the peculiar dependence is not connected to excitation-induced nonradiative dynamics, associated with several kinds of incoherent scatterings and/or sample heating. Rather, the observed feature originates from Rabi oscillations, i.e., occupation exchanges of a two-level system through coherent excitations. 

In the case of an ideal two-level system, Rabi oscillations are expected to follow, 
\begin{equation}
I(t) \propto \sin^2\left[ \Theta(t)/2 \right], \quad  \Theta(t)=\frac{\mathcal{D}}{\hbar} \int^t_{-\infty} E(t')dt',
\label{Eq_rabi}
\end{equation} 
where $\mathcal{D}$ and $E(t)$ are a dipole moment and a field envelope, respectively. For short pulsed excitation, we measure the emission intensity after so that $I(t\to\infty)$. The parameter $\Theta(t\to\infty)\equiv \Theta$ is known as pulse area. According to Eq.~\ref{Eq_rabi}, the oscillation behaves as a sinusoidal curve, while our result suggests the presence of other components being superimposed on the oscillation. We attribute these components to the emissions of the excited states of other QDs, and/or the spectral continuum typically appearing at high excitation, caused by a coupling between QDs and an two-dimensional layer. We take account of this effect by simply introducing a linearly-dependent term to Eq.~\ref{Eq_rabi}, that is, 
\begin{equation}
I(t\to\infty) \propto \sin^2\left( \Theta/2 \right)+a\Theta^2.  
\label{Eq_rabi_2}
\end{equation}
Note that the term $\Theta^2$ is proportional to excitation intensity. This expression well reproduces our result as shown in Fig.~\ref{fig_rabi}.

We can determine the dipole moment of our QD by analyzing Rabi oscillations: We assume a hyperbolic secant-squared pulse, 
$E(t)=E_{0}\,\textrm{sech} (t/\tau_{p})$, 
where $1.763\times \tau_p$ gives FWHM in the field envelope. The pulse area in this case is expressed by 
$\Theta=\mathcal{D} E_{0}  \pi \tau_{p}/\hbar$. 
According to the fit, we find $\Theta=\pi$ at $16\,(\pm 2) \sqrt{\textrm{W/cm}^2}$, corresponding to pulse energy density $P=0.21\,(\pm 0.03)$ $\mu$J/cm$^2$/pulse. 
The field amplitude $E_0$ is obtained by the expression of electro-magnetic energy inside a pulse packet, i.e.,  
$P=cn\epsilon_{0}\int|E(t')|^{2}dt'=cn\epsilon_{0}E_{0}^2 2\tau_{p}$, 
where $c$ and $n$ are the speed of light in a vacuum, and a refractive index, respectively. 

The pulse temporal width $\tau_p$ in the above treatment is defined for the case of Fourier-transform limited pulses, but not for actual pulses. For general chirped pulses, the time $\tau_p$ agrees well with the FWHM of the first-order correlation function of excitation pulses $(g^{(1)}(\tau))$. So we evaluated the correlation function using a Michelson-type interferometer, and determined it to be $\tau_p= 2.4$ ps. With $n= 3.6$ for GaAs, we eventually obtained the dipole moment of our QD, $\mathcal{D}= 23\,(\pm 4)$ debye. 

Since the spatial extent of QDs is much smaller than the wavelength of light, the spontaneous emission is expected to follow the atomic two-level description. The decay rate in a material is, therefore, given by the Einstein A coefficient, \cite{Meystre}
\begin{equation}
\Gamma_A=n\frac{\mu}{\mu_0}\frac{\omega^3\mathcal{D}^2}{3\pi\epsilon_{0} \hbar c^3}, 
\label{rate}
\end{equation}
where $\omega$ is the angular frequency of light. By substituting $\mathcal{D} = 23\,(\pm4)$ debye and other material parameters into Eq.~\ref{rate}, we find $\Gamma_A^{-1}= 590\,\binom{+280}{-160}$~ps. This value is, however, significantly larger than the PL decay time that was observed in PL transients (220 ps). The reduction in PL decay time compared with the ideal value may have arisen from the effect of nonradiative relaxation involved in our QDs. Further studies are needed to clarify the origin of the rapid PL relaxation.

\begin{figure}
\includegraphics[angle=-90,width=7cm,clip]{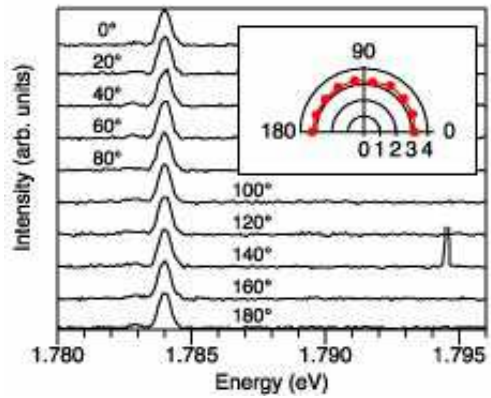} 
\caption{(color online) Polarized PL spectra of a single GaAs QD with varying angle of linear polarization. The energy of excitation was above the (Al,Ga)As barrier. 
The inset shows the integrated intensities as a function of polarization angle.}
\label{polar} 
\end{figure}

Finally, we discuss on the spin property in our QDs: We observed the emission signals whose polarization was perpendicular to that of the excitation beams. The presence of such emissions suggests the time scale of depolarization (i.e., spin coherence relaxation) being smaller than that of recombination. A possible origin for the fast spin decoherence is a lowering of the spherical symmetry due to shape anisotropy depending on each dot. Note that our GaAs/(Al, Ga)As QDs are free from strain, and the fine energy split is considered to be small, or possibly absent. In Fig.~4 we show polarized PL spectra with varying angle of polarization, indicating the spectral shift being far below resolution ($< 50$ $\mu$eV). The situation is contrast to Stranski-Krastanow grown QDs \cite{Seguin_PRL05}. Moreover, the Fourier-transform measurement confirmed the exciton PL consisting of a single spectral component \cite{kuroda_apl06}. As a result, spin-related sublevels are mostly degenerate in our QDs, and energetic relaxation is not involved in our present excitation/detection scheme.

In summary, we demonstrated the read out of excitonic qubits in 
GaAs QDs by observing time-resolved spontaneous emissions. Weak emissions from 
QDs were successfully captured by suppressing the scattering of excitation beams. Excitonic Rabi oscillations were observed in the excitation-intensity dependence of spontaneous emissions. We believe that our method provides a simple and effective technique to read out the final state of an excitonic qubit. Application of this technique to high-yield, single-photon emitting devices could enable single-shot detection in single exciton qubits. 

The authors would like to thank T.~Ochiai, J.~Inoue, M.~Yamagiwa, and Professor M.~Kawabe for their fruitful discussions.  



\begin{thebibliography}{99}
\bibitem{EHW}
	J.~M.~Elzerman, R.~Hanson, L.~H.~Willems~van~Beveren, B.~Witkamp, 
	L.~M.~K. Vandersypen, and L.~P.~Kouwenhoven, 
	Nature \textbf{430}, 431 (2004).
	
\bibitem{Kamada}
	H. Kamada, H. Gotoh, J. Temmyo, T. Takagahara, and H. Ando, 
	Phys. Rev. Lett. \textbf{87}, 246401 (2001). 

\bibitem{Htoon}
	H. Htoon, T. Takagahara, D. Kulik, O. Baklenov, A. L. Holmes,  and K. Shih, 
	Phys. Rev. Lett. \textbf{88}, 087401 (2002). 
	
\bibitem{Muller}
	A. Muller, Q. Q. Wang, P. Bianucci, C. K. Shih, and Q. K. Xue, 
	Appl. Phys. Lett. \textbf{84}, 981 (2004); 
	\textit{ibid.} \textbf{87}, 031904 (2005).
	
\bibitem{Zrenner}
	A. Zrenner, E. Beham, S. Stufler, F. Findeis, M. Bichler, and G. Abstreiter, 
	Nature \textbf{418}, 612 (2002).
	
\bibitem{Stufler}
	S. Stufler, P. Ester, A. Zrenner, and M. Bichler, 
	Phys. Rev. B \textbf{72}, 121302(R) (2005). 
	 
\bibitem{Stievater}
	T. H. Stievater, X. Li, D. G. Steel, D. Gammon, D. S. Katzer, D. Park, 
	C. Piermarocchi and L. J. Sham, 
	Phys. Rev. Lett. \textbf{87}, 133603 (2001). 

\bibitem{Li}
	X. Li, Y. Wu, D. G. Steel, D. Gammon, T. H. Stievater, D. S. Katzer, D. Park, 
	C. Piermarocchi and L. J. Sham, 
	Science \textbf{301}, 809 (2003).

\bibitem{Borri}
	P. Borri, W. Langbein, S. Schneider, U. Woggon, R. L. Sellin, D. Ouyang, 
	and D. Bimberg, 
	Phys. Rev. B \textbf{66}, 081306(R) (2002).

\bibitem{Mitsumori}
	Y. Mitsumori, A. Hasegawa, M. Sasaki, H. Maruki, and F. Minami, 
	Phys. Rev. B \textbf{71}, 233305 (2005).

\bibitem{paillard_apl00}
	M.~Paillard, X.~Marie, E.~Vanelle, T.~Amand, V.~K.~Kalevich, A.~R.~Kovsh, 
	A.~E.~Zhukov, and V.~M.~Ustinov, 
	Appl. Phys. Lett. \textbf{76}, 76 (2000). 

\bibitem{KTC91}
	N. Koguchi, S. Takahashi, and T. Chikyo, 
	J. Cryst. Growth \textbf{111}, 688 (1991). 

\bibitem{Watanabe1}
	K. Watanabe, N. Koguchi, and Y. Gotoh, 
	Jpn. J. Appl. Phys. Part 2, \textbf{39}, L79(2000).

\bibitem{Watanabe2}
	K. Watanabe, S. Tsukamoto, Y. Gotoh, and N. Koguchi, 
	J. Cryst. Growth \textbf{227-228},1073 (2001). 

\bibitem{kuroda_prb02}
	T. Kuroda, S. Sanguinetti, M.Gurioli, K. Watanabe, F. Minami, and N. Koguchi, 
	Phys. Rev. B \textbf{66}, 121302 (2002). 

\bibitem{kuroda_apl06}
	K. Kuroda, T. Kuroda, K. Watanabe, K. Sakoda, N. Koguchi, and G. Kido, 
	Appl. Phys. Lett. \textbf{88}, 124101 (2006). 


\bibitem{Meystre}
	See, e.g., P. Meystre and M. Sargent III, 
	{\it Elements of Quantum Optics}, 2nd ed. (Springer-Verlag, Berlin, 1991).

\bibitem{Seguin_PRL05}
	R. Seguin, A. Schliwa, S. Rodt, K. P\"otschke, U. W. Pohl, and D. Bimberg, 
	Phys. Rev. Lett. \textbf{95}, 257402 (2005).

\end{thebibliography}
\end{document}